\documentclass[11pt,prd,onecolumn,showpacs,amsmath,amssymb,aps,floats,floatfix]{article}
\usepackage{float}
\usepackage[flushleft]{threeparttable}
\usepackage[normalem]{ulem}  
\usepackage[section]{placeins}
\usepackage{authblk}

\usepackage{aas_macros}
\usepackage[sort,numbers]{natbib}
\usepackage{gensymb}
\usepackage{hyperref}
\usepackage{comment}
\usepackage{dcolumn}
\usepackage{bm}
\usepackage{amsmath}
\usepackage{amsfonts}
\usepackage{amssymb}
\usepackage{latexsym}
\usepackage{slashed} 
\usepackage{pstricks}
\usepackage{indentfirst}
\usepackage{mathrsfs}
\usepackage{multirow}
\usepackage{epsfig,psfrag}
\usepackage{subfigure}
\usepackage{setspace} 

\usepackage{amsmath}
\usepackage{footnote}

\usepackage{indentfirst}

\usepackage{graphicx}
\usepackage{epsfig,subfigure,psfrag}
\usepackage{mathrsfs}
\usepackage{slashed} 
\def \be {\begin{equation}}
\def \ee {\end{equation}}
\def \ba {\begin{eqnarray}}
\def \ea {\end{eqnarray}}
\usepackage{color}
\definecolor{darkgreen}{rgb}{0,0.3,0}
\definecolor{darkblue}{rgb}{0,0,0.2}
\definecolor{darkred}{rgb}{0.5,0,0}

\definecolor{lightgrey}{rgb}{0.8,0.8,0.8}
\definecolor{black}{rgb}{0,0,0}

\def\AU{\mathrm{AU}}
\def\MeV{\mathrm{MeV}}
\def\GV{\mathrm{GV}}
\def\nT{\mathrm{nT}}

\newcommand {\dbar}{{d\kern-.22em\lower-.73ex\hbox{-}}}

\def\exp{\mathrm{exp}}
\def\sr{\mathrm{sr}}

\def\GV{\mathrm{GV}}

\def\cm{\mathrm{cm}} 
\def\km{\mathrm{km}} 
\def\s{\mathrm{s}} 
\def\GeV{\mathrm{GeV}} 
\def\MeV{\mathrm{MeV}} 

\def\m{\mathrm{m}}
\makeindex

\begin{document}
\title{Time dependent solar modulation of cosmic rays from solar minimum to solar maximum}
\author[1,2]{Bing-Bing Wang}
\author[1,2]{Xiao-Jun Bi}
\author[1]{Kung Fang}
\author[1]{Su-Jie Lin}
\author[1]{Peng-Fei Yin}
\affil[1]{Key Laboratory of Particle Astrophysics, Institute of High Energy Physics,
Chinese Academy of Sciences, Beijing 100049, China}
\affil[2]{School of Physical Sciences, University of Chinese
Academy of Sciences, Beijing 100049, China}
\maketitle


\begin{abstract}
We study the time-dependent modulation effect and derive the local interstellar spectra (LIS) for the cosmic ray (CR) proton, helium, boron and carbon. 
A two-dimensional modulation model including the variation of the interplanetary environment with time is adopted to describe modulation process.
The propagation equation of CRs in the heliosphere is numerically solved by the package Solarprop. 
We derive the LIS by fitting the latest results of several experiments, including Voyager 1, PAMELA, BESS-POLARII and ACE, during low solar activity periods. 
We further study the modulation in the polarity reversal periods with the PAMELA proton data. 
We find that the rigidity dependence of the diffusion coefficient is critical to explain the modulation effect during reversal periods. 
Our results also indicate a power law relation between the diffusion coefficient and the magnitude of the heliospheric magnetic field (HMF) at the Earth.
\end{abstract}

\section{Introduction}
After accelerated in sources, CRs are injected and propagated in the Galactic interstellar space. 
When entering the heliosphere, the intensities of CRs at low energies are significantly affected by several local
effects, such as the interactions with the outward solar wind with an embedded magnetic field \cite{1965P&SS...13....9P}. 
Therefore, the observed CR spectra are modulated with the solar activity cycle, and are different from those outside the heliosphere, namely the local interstellar spectra. 
The study of the solar  modulation is essential, as it is indispensable for reproducing the low energy ($<30\,\GeV$) LIS and can also help us to understand the physical process in the CR-heliosphere interaction.
The LIS are critical to determine the injection information and the propagation model of Galactic CRs
\cite{2016PhRvD..94l3019K,2017PhRvD..95h3007Y,2017PhRvD..96j3005T,2018PhRvD..97b3015N,2018arXiv180904905W,2018arXiv181003141Y,2019SCPMA..6249511Y}, which are also closely related to some new physics studies, such as the indirect detection of dark matter
\cite{yin2009pamela,2013JCAP...09..031F,2014JCAP...12..045C,2015APh....60....1Y,2015PhRvD..91f3508L,2015arXiv150407230L,
2018SCPMA..61j1004W}.

The current data of CR experiments provide unprecedentedly good opportunities
for the research of solar modulation. 
The Voyager 1, which has crossed the boundary of heliosphere (heliopause) since August 2012, can give direct measurements of CR LIS from a few to hundreds MeV/nucleon
\cite{2013Sci...341..150S,2013Sci...341.1489G,2016ApJ...831...18C}. 
Among the experiments, the PAMELA experiment \cite{Picozza:2006nm,Boezio:2009zz} is particularly compelling for the study of solar modulation \footnote{Investigating the solar modulation with AMS-02 data is in preparation.}. 
The PAMELA collaboration has published eight years of CR data (2006/07-2014/02), continuously recording the variation of the CR proton spectrum from the late declining phase of solar cycle 23 to the maximum phase of solar cycle 24 \cite{2013ApJ...765...91A,2018arXiv180110310A}. 
Moreover, PAMELA performs precise measurements of the proton spectrum in a wide energy range of $80\,\MeV$ $-$ $50\, \GeV$, partly overlapping with the energy range of Voyager 1.
Then combining the results of PAMELA and Voyager 1, we can give good constraints to the model of solar modulation and obtain a reasonable CR proton LIS.

The most widely used model of solar modulation is the force field approximation \cite{1968ApJ...154.1011G}.
It is oversimplified to deal with all the current precise data. The more detailed models have been developed to interpret the observations \cite{2016ApJ...829....8C,2016PhRvD..93d3016C,2017JGRA..12210964G,Kappl:2015hxv,2017ICRC...35...24V,2018AdSpR..62.2859B,2017ApJ...846...56Q,Potgieter:2013cwj,Potgieter:2017zat,Orcinha:2018tvx,2018shin.confE..92C,2018PhRvL.121y1104T,2019ApJ...871..253C,2019ApJ...873...70A}.
In this work, we consider a time-dependent 2D modulation model including the diffusion, convection, drift and adiabatic energy loss processes, to study the modulation effect over different solar activity periods and derive the CR LIS. 
Based on the public code Solarprop \footnote{\url{http://www.th.physik.uni-bonn.de/nilles/people/kappl/}} \cite{2016CoPhC.207..386K}, we numerically solve the propagation equation of CRs in the heliosphere with stochastic differential equation approach. 
The typical parameters related to the interplanetary medium environment, such as the magnitude of the heliospheric magnetic field (HMF), the solar wind speed, and the tilt angle
of the heliospheric current sheet (HCS), are taken from the observations. 
The scale factor of the diffusion coefficient is set to be a time-dependent free parameter to accommodate the observations.

In order to obtain the proton LIS in the form of cubic spline interpolation, we simultaneously fit the Voyager 1 data and the PAMELA data \cite{2016ApJ...831...18C,2013ApJ...765...91A,2018ApJ...854L...2M} in some low solar
activity periods. 
Then we attempt to explain the PAMELA results in the polarity reversal periods, which is more challenging compared with the case of less active periods.
Besides, ACE and BESS experiments provide time-dependent spectra for CR
nuclei \cite{2006AdSpR..38.1558D,2013ApJ...770..117L,2007APh....28..154S,2016ApJ...822...65A}, and Voyager 1 also measures the low energy LIS of CR nuclei \cite{2016ApJ...831...18C}. We then derive the LIS of helium, boron and carbon in the same framework with the study
of proton.

The paper is organized as follows. In Section ~\ref{sec:envi}, we present the important ingredients describing the heliosphere environment. In Section ~\ref{sec:prop}, we briefly describe the dominant physical mechanisms in the solar modulation effect.
In Section ~\ref{sec:proton}, we drive the proton LIS using the PAMELA data and compare the calculated spectra with the PAMELA results over different solar activity periods. We also give an empirical relation between diffusion coefficient and heliospheric magnetic field strength at the Earth.
In Section ~\ref{sec:nuclei} we present the LIS of helium, boron and carbon.
Finally, we give the summary in Section ~\ref{sec:conclusion}.

\section{The global characteristics of the heliospheric environment}\label{sec:envi}
The CR propagation in the heliosphere is affected by the solar activity. 
The main factors that affect the CR transportation, namely solar wind speed, the magnitude and orientation of heliospheric magnetic field and the inclination between the heliospheric current sheet and the equatorial plane named tilt angle, are all correlated with the solar activity.

The solar wind speed is variable in both latitude and time. During the solar minimum, the typical solar wind speed at the low latitude is about $400\,\km /\s$, while it increases by almost a factor of two at the high latitude \cite{2000JGR...10510419M}. With the increase of solar activity, the boundary of slow and fast solar wind rises rapidly \cite{2000GeoRL..27..149W}.
The solar wind speed is described as \cite{Bobik:2011ig}
\begin{equation}
\begin{aligned}[c]
	V_{sw} = \begin{cases}
		V_{max}, & \theta \leq 30\degree \,or\, \theta \geq 150\degree . \\
		V_{min}(1+\lvert \cos \theta \rvert), & 30\degree < \theta < 150\degree.
\end{cases}
\end{aligned}
\end{equation}
where $V_{max}$ is taken to be $760\,\km/\s$ and $V_{min}$ is the observation value near the Earth. For simplicity, we take the latitude average value as an approximation in this work.

The solar wind carries the Sun's magnetic field into the interplanetary space and forms the heliospheric magetic field (HMF) with an Archimedean spiral structure given by \cite{1958ApJ...128..664P}:
\begin{equation}
\vec{B} = A\frac{B_0}{r^2} (\vec{e}_r - \tan \psi \vec{e}_{\phi})(1-2H(\theta-\theta_{ns})),
\end{equation}
where $A=+$ ($A=-$) indicates the magnetic field polarity for the solar magnetic field lines pointing outward (inward) in the northern hemisphere and inward (outward) in the southern hemisphere, $r$ is the distance from the Sun, $B_0 \sqrt{(1+\tan^2\psi(r=1\AU,\theta=\pi/2))}$ is the magnetic filed magnitude at the Earth, $H$ is Heaviside function, $\theta$ is the polar angle and $V_{sw}$ is the solar wind speed. The spiral angle $\psi$ is defined as
$\tan{\psi}=\frac{\Omega r \sin \theta}{V_{sw}}$, where $\Omega = 2.866\cdot 10^{-6} ~ \rm{rad} /\rm{s}$ is the rotation speed of the Sun. $\theta_{ns}$ determines the position of heliospheric current sheet (HCS), which divides the heliosphere into two regions with opposite polarities. The structure of the HCS is parameterized by its tilt angle and is well related to the solar activity.
Some modifications of the Parker HMF have been proposed in \cite{1989GeoRL..16....1J,1991ApJ...370..435S,1996JGR...10115547F}, but remain inconclusive. We adopt the standard Parker HMF model in this work.

\begin{figure}[!hbt]	
    \centering
	\includegraphics[width=.9\textwidth]{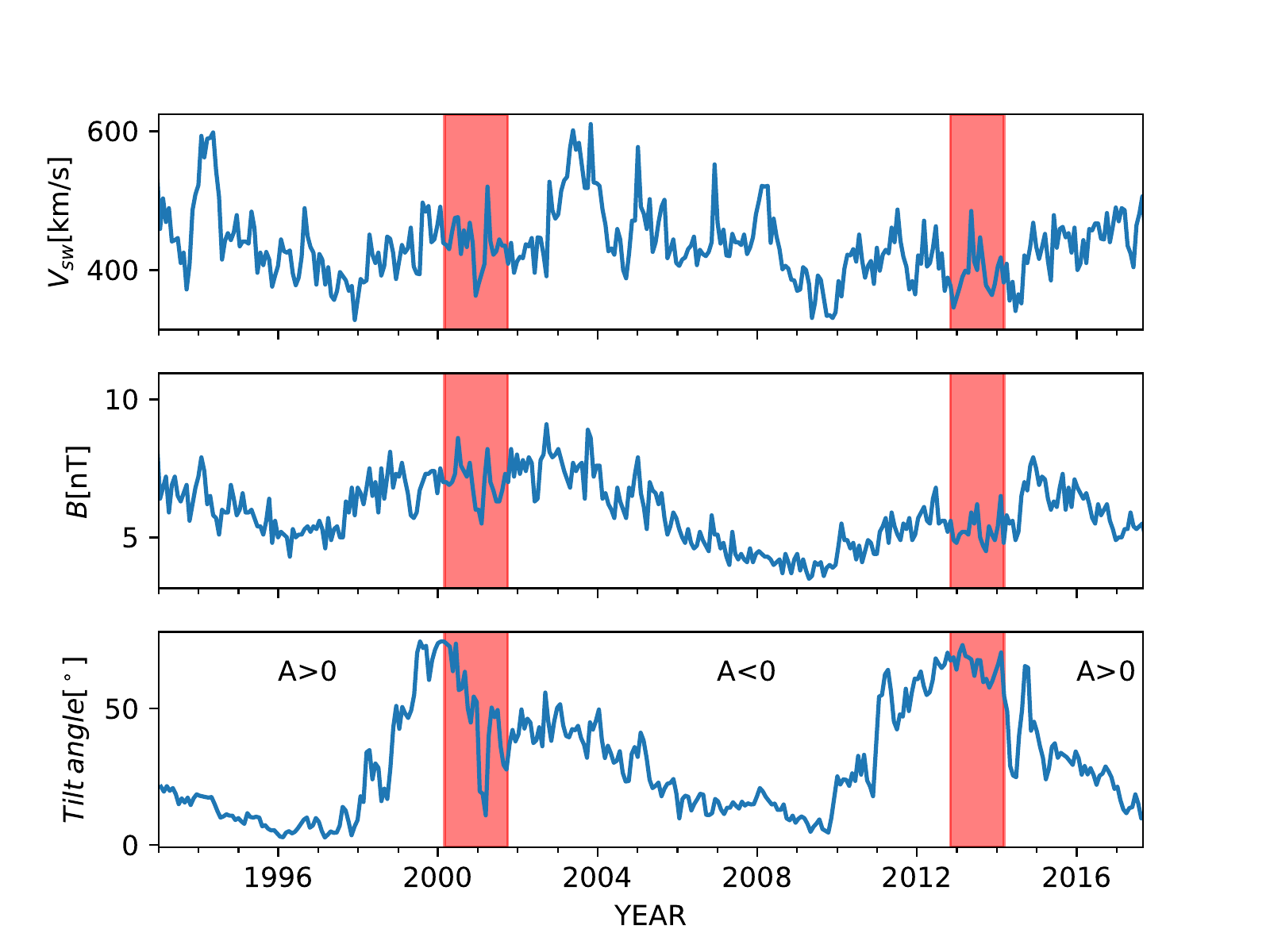}
    \caption{Time profiles of the input interplanetary parameters in the solar modulation model. The top and middle panels show the solar wind speed and the magnetic field strength for each Carrington rotation taken from the OMNI website interface (\url{omniweb.gsfc.nasa.gov}), respectively. The bottom panel represents the tilt angle of HCS taken from the WSO website (\url{wso.stanford.edu}) with the ``new'' model. The red boxes represent the estimated polarity reversal periods \cite{2003ApJ...598L..63G,2015ApJ...798..114S}.}
\label{fg:BVA}
\end{figure}

In Figure ~\ref{fg:BVA} we show the time profile of the parameters characterizing the global heliospheric environment in our model. The first and second panels show the averaged solar wind speed ($V_{sw}$) and HMF magnitude at $1 \,\AU$ ($B_E$) using the data from OMNI website interface\footnote{\url{omniweb.gsfc.nasa.gov}} for each Carrington rotation (about 27.28 days), respectively. The third panel illustrates the variation of the tilt angle ($\alpha$) taken from the WSO website
\footnote{\url{wso.stanford.edu}} with the ``new'' model. The estimated periods of changeover of the solar magnetic polarity \cite{2003ApJ...598L..63G,2015ApJ...798..114S} are represented by the red bands.
During the time period considered in this work, the solar wind speed and the magnetic field magnitude near the Earth vary in the ranges of 328--610 $\km/\s$ and 3.1--9.1 $\nT$, respectively. 
The tilt angle have a obvious variation from $4.5^\circ$ to $74.5^\circ$. 
Both the HMF strength and tilt angle show a clear nearly 11-year cycle.

As shown in Figure ~\ref{fg:BVA}, the description for the heliosphere environment with fixed parameters is not realistic.
In the original Solarprop, not only the magnetic field strength but also the tilt angle change with time. We further extend the code to allow the variation of solar wind speed.
In this work, we study the solar modulation effect including a smooth time correlation to the solar activity.
As a valid approximation, the input parameters are averaged in 7--12 months, which denote the time scale of the solar wind propagation from the Sun to the modulation boundary.

\section{The CR propagation in the heliosphere}\label{sec:prop}
The CR propagation within heliosphere is dominantly affected by four effects, including the diffusion resulted from scattering magnetic irregularities, the convection caused by outward solar wind, the drift induced by the irregularity of the global heliosphere magnetic field, and the adiabatic energy loss \cite{2013LRSP...10....3P}.
The modulation effect in the heliosheath is neglected here. For the discussion of this possible effect, we refer the reader to \cite{2011ApJ...735..128S,2012cosp...39.1714S,2013ApJ...765L..18S,2014ApJ...782...24K}. The review of the solar modulation can be found in \cite{2006SSRv..125...81H,2013LRSP...10....3P}.

The CR propagation within heliosphere can be described by the Parker transport equation \cite{1965P&SS...13....9P}:
\begin{equation}
\frac{\partial f}{\partial t} = -(\vec{V}_{sw} +\vec{V}_{drift})\cdot \nabla f + \nabla \cdot[K^s\cdot \nabla f] + \frac{\nabla \cdot \vec{V}_{sw}}{3} \frac{\partial{f}}{{\partial\ln p}} ,
\end{equation}
where $f(\vec{r},p,t)$ is the omni-directional distribution function, $\vec{r}$ is the position in a heliocentric spherical coordinate system, $p$ is the particle momentum, $\vec{V}_{sw}$ is the solar wind speed, $\vec {V}_{drift}$ is the drift speed, and $K^s$ is the symmetric part of the diffusion tensor. The differential intensity related to the distribution function is given by $I = p^2f$.

It is far from completely understanding the parameterization of the diffusion tensor \cite{2010Ap&SS.325...99S,2015ApJ...814..152E,2017ApJ...846...56Q,2018ApJ...856...94Z}.
In this work, a simple spatial and rigidity dependence of the parallel diffusion coefficient $K_{\parallel}$ is adopted as the following form \cite{2016CoPhC.207..386K}
\begin{equation}
	\begin{aligned}[c]
K_{\parallel} = & \begin{cases}
		& \frac{1}{30} k \beta \frac{B_E}{B} , R<0.1\,\GV \\
		& \frac{1}{3} k \beta R \frac{B_E}{B}, R \geq 0.1\,\GV
	\end{cases}
	\end{aligned}
\end{equation}
where $k \equiv k_0 \cdot 3.6 \times 10^{22} \cm^{2}/\s$ is a scale factor, which describes the time dependence of the diffusion coefficient and reflects the variability of interplanetary medium properties, $B_E$ is the strength of HMF at the Earth, and $B$ is the strength of HMF at the particle position.
This rigidity dependence is suggested for modulation during the solar minimum in \cite{1968CaJPS..46..950J} based on the quasilinear theory \cite{1966ApJ...146..480J}.
For particle rigidity above a threshold value the linear rigidity dependence of diffusion coefficient is commonly adopted in many works \cite{1968ApJ...154.1011G,1987A&A...184..119P,1994ApJ...423..817P,2001SSRv...97..373W,2011ApJ...735...83S,2012ApJ...745..132B,2018arXiv181108909T}.
The diffusion coefficient perpendicular to the large scale HMF $K_{\perp}$ is taken to be $K_{\perp} = 0.02K_{\parallel}$ according to the test particle simulation \cite{1999ApJ...520..204G}. 
This form of $K_{\perp}$ is widely adopted in literatures \cite{2002JGRA..107.1221F,Potgieter:2013cwj}.

The drift effect leads to a charge-sign dependence and a 22-year cycle in the solar modulation effect \cite{1976JGR....81.2082L,1977ApJ...213..861J,1981ApJ...243.1115J}.
The drift speed from the gradient and curvature of the HMF is written as $V_{gc} = q \frac{\beta R}{3} \, \nabla \times \frac{\vec{B}}{B^2}$ \cite{1977ApJ...213..861J}.
The HCS drift is caused by the change of the field direction at the crossing of the HCS.
In this study, we describe the HCS drift following \cite{1989ApJ...339..501B}, where a thick, symmetric transition region determined by the tilt angle is used to simulate a wavy neutral sheet. 
Combining the $V_{gc}$ and effective wavy neutral sheet drift speed $V_{ns}^{w}$ together show that the drift speed is divergence-free in the region of $\pi/2-\alpha-\theta_{\triangle} < \theta <\pi/2 + \alpha +\theta_{\triangle}$, where $\delta_{\triangle} \approx \frac{2RV_{sw}}{AB_0\Omega \cos{\alpha}}$.
The $V_{ns}^{w}$ is given by
\begin{equation}
   \begin{aligned}
		\vec{V}_{ns}^{w}  = & \begin{cases}
		qA \frac{v \theta_{\triangle} \cos(\alpha)}{6\sin(\alpha+\theta_{\triangle})} \vec{e}_{r} ,& \pi/2-\alpha-\theta_{\triangle} < \theta <\pi/2 + \alpha +\theta_{\triangle} \\
      0, & else 
	\end{cases}
   \end{aligned}
\end{equation}
where $q$ is the charge sign and $v$ is the particle speed. 
The product of $qA$ determines the drift direction.
During the $A<0$ cycle the positive charge particles drift inwards mainly through the HCS near the equatorial regions. Otherwise, during the $A>0$ cycle, they mainly drift inwards from polar regions.

In this work, we use the public monte carlo code Solarprop \cite{2016CoPhC.207..386K} to numerically solve the transport equation. The computation is based on the equivalence of a set of stochastic differential equations to the Parker equation. For the details of the numerical method, we refer the reader to \cite{1998GeoRL..25.2353Y,1999ApJ...513..409Z,2010JGRA..11512107P,2016CoPhC.207..386K}.
We take the termination shock as the modulation boundary and assume that it is $100 \,\AU$ from the Sun.
In our default calculation, the only free parameter is the time dependent scale factor of the diffusion coefficient $k_0$. Other input parameters are obtained from observations, such as the solar wind speed, the magnitude of magnetic field and the HCS tilt angle.

\section{The solar modulation for proton}\label{sec:proton}
The PAMELA experiment performed a systematic measurement of the CR proton spectrum in the period 2006-2014 from the late declining phase of solar cycle 23 to the maximum of cycle 24.
The detailed comparison between the calculated energy spectra and the observations for different solar activity levels can improve our understanding about the modulation process.

\subsection{The local interstellar spectrum for proton}
The Voyager 1 crossed the heliopause on the August 2012 and provided the CR LIS at very low energies. 
Additionally, the CR spectra measured by PAMELA and AMS-02 with rigidity above a few tens of $\GV$ are not affected by the modulation.
However, until now, there have been no experiments to measure the LIS in the gap.
To derive the LIS we parameterize the LIS with the cubic spline interpolation method \cite{2016A&A...591A..94G,2018ApJ...863..119Z}.
We obtain the proton LIS by fitting the calculated spectra to the Voyager 1 observation and a series of PAMELA data. 
The GNU Scientific Library (GSL)\footnote{\url{https://www.gnu.org/software/gsl/}} is used to perform the least-sqaures fitting.
In order to avoid the influence of the polarity reversal occurred in the late of 2012, the PAMELA sample data are chosen between Jul. 2006 to Feb. 2012.
For considerable saving in computing time we use 12 sets of data to construct the LIS. 
The corresponding Carrington rotation number of the data used in the fit are 2045, 2052, 2058, 2064, 2070, 2076, 2082, 2088, 2093, 2107, 2114, 2121.

\begin{figure}[!hbt]
	\centering
	\includegraphics[width=0.9\textwidth]{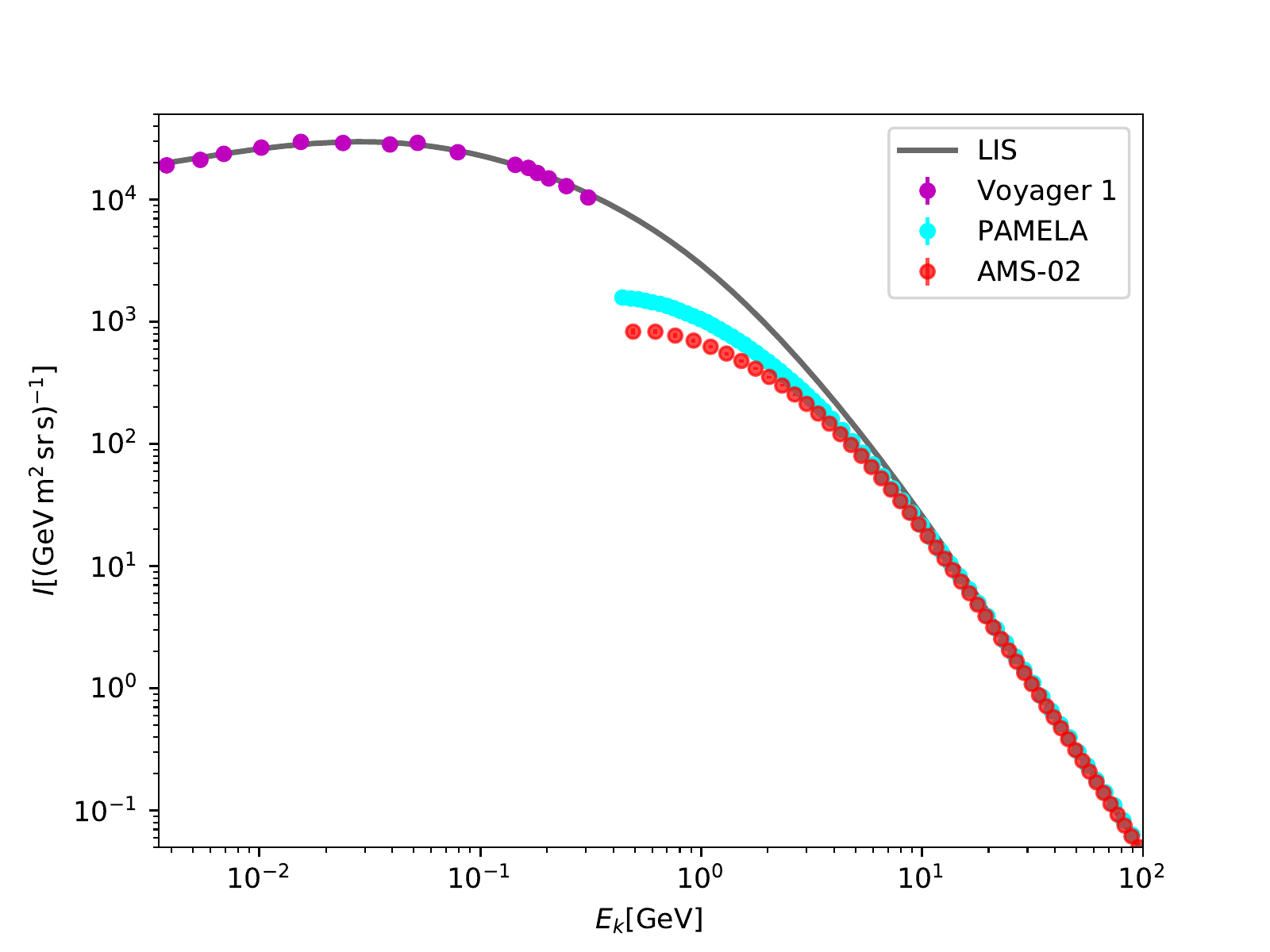}
	\caption{Comparison of the proton LIS to the observations. The grey curve represents the proton LIS. The purple, cyan and red dots represent the Voyager 1, PAMELA and AMS-02 data, respectively. }
	\label{fg:lis}
\end{figure}

\begin{table}[!htb]
	\caption{The parameterization of proton LIS with cubic spline interpolation.} 	
	\centering
	\begin{tabular}{lcccccccl}\hline \hline
	
		$log(E_k/\GeV)$	 			&-2.42	 &-1.41	&-0.50	&0		&0.50		&1.00		&1.50		&2.00 \\
		$log(I/(\GeV\m^2\sr\,\s))$	&4.3003			 &4.4676			&4.0396			&3.4675		&2.5702			&1.4101			&0.0685			&-1.3465		\\
	\hline
	\end{tabular}\label{tab:lis}
\end{table}

The knots and the corresponding proton intensities are listed in Table ~\ref{tab:lis}.
Figure ~\ref{fg:lis} shows the obtained proton LIS and the experimental results, including the Voyager 1, PAMELA, and AMS-02 data \cite{2016ApJ...831...18C,2011Sci...332...69A,2015PhRvL.114q1103A}.
The LIS agrees with the proton flux measured by Voyager 1 outside the heliosphere at low energies below $300\, \MeV$, and is consistent with the data measured by PAMELA and AMS-02 at high energies above a few $10\,\GeV$. 
The LIS shows that it is less affected by the solar modulation for proton above $10\,\GeV$.

\subsection{Comparing the calculated proton spectra with the observations before the polarity reversal}

After deriving the proton LIS, we can calculate the modulated proton spectra in different periods, and compare them to the spectra observed by PAMELA.
We firstly focus on the modulation before the polarity reversal. The northern and southern polar fields reversed in 2012 November and 2014 March, respectively \cite{2015ApJ...798..114S}.
In every period, the diffusion coefficient is adjusted to reproduce the observed PAMELA spectrum in the range of 0.08--40 $\GeV$.

\begin{figure}[!hbt]
	\centering
	\begin{tabular}{@{}cc@{}}
	\includegraphics[width=0.55\textwidth]{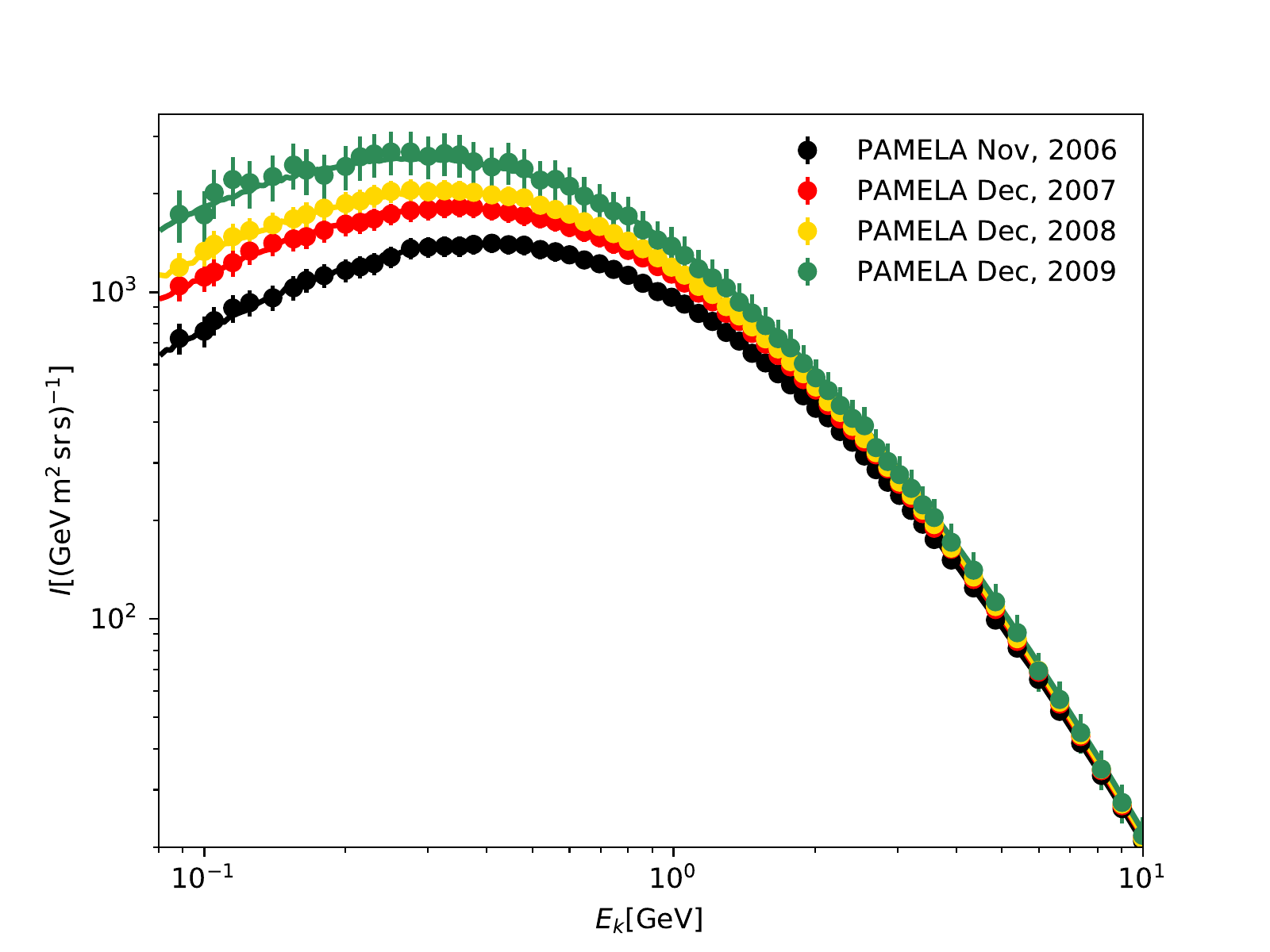}
	\includegraphics[width=0.55\textwidth]{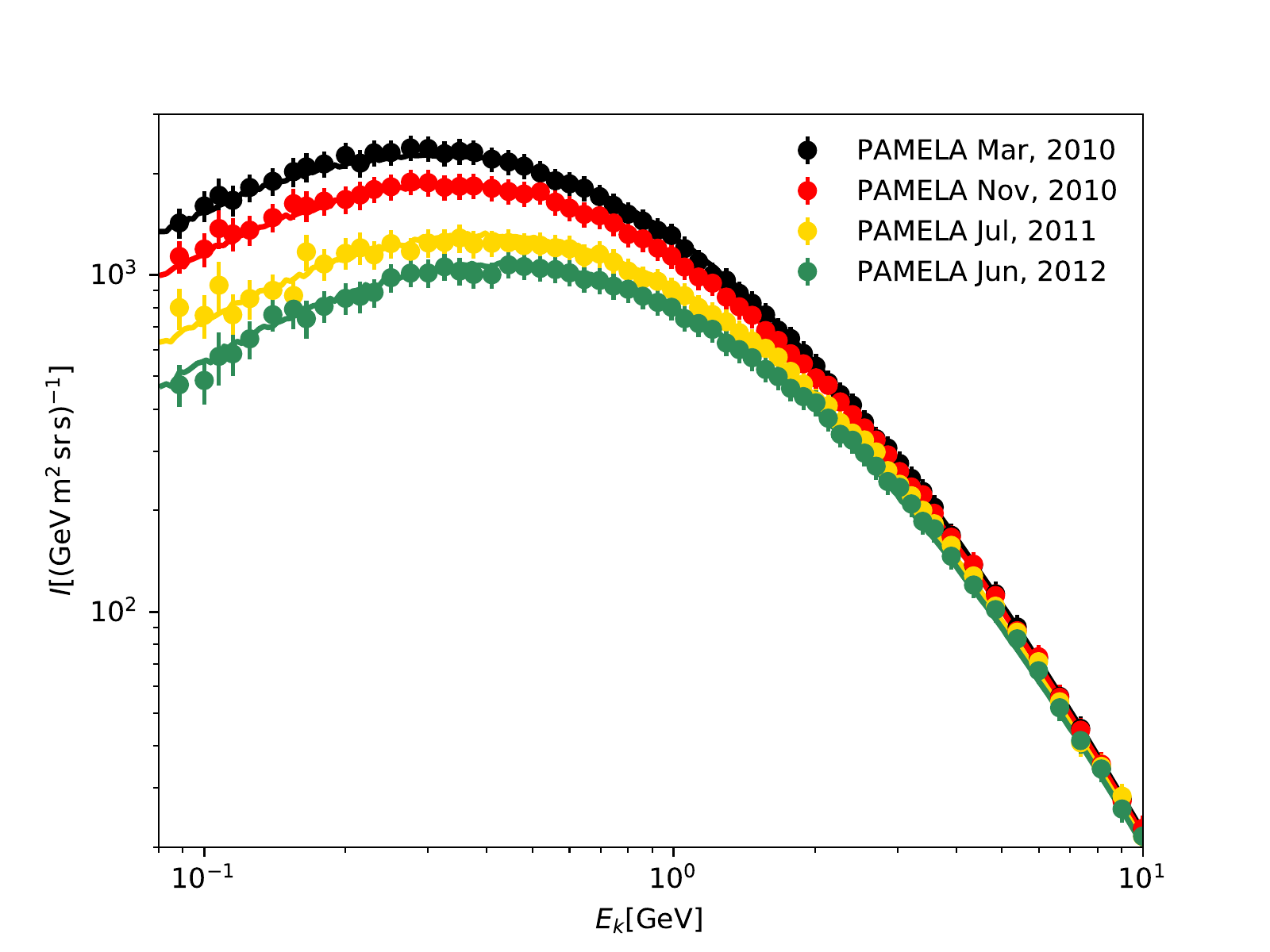}
	\end{tabular}
	\caption{The calculated proton spectra (solid lines) are compared to a selection of the PAMELA data (dots) in 2006-2009 (left panel) and 2010-2012 (right panel). }
	\label{fg:proton_pamela}
\end{figure}

\begin{figure}[!hbt]
	\centering
	\includegraphics[width=0.9\textwidth]{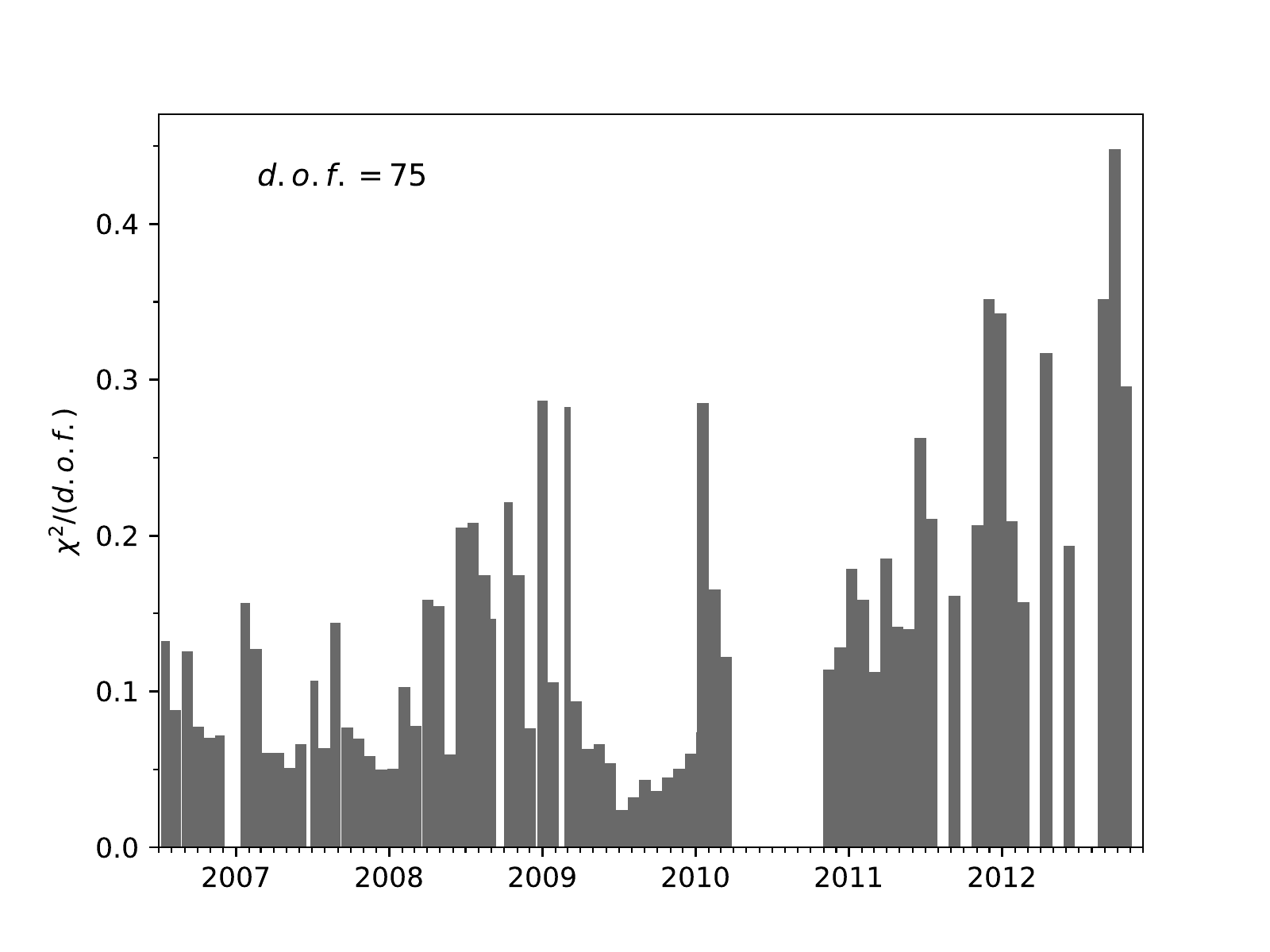}
	\caption{The time profile of reduced-$\chi^2$ from the fit to the PAMELA proton data during 2006/07 to 2012/10.}
	\label{fg:chi1}
\end{figure}

In Figure ~\ref{fg:proton_pamela}, we show that our calculated proton spectra are consistent with the corresponding PAMELA data over a wide energy range in 8 different periods. 
The time evolution of proton flux is close related to the solar activity.
The proton flux gradually increased from 2006 to 2009 until reaching the maximum value during 2009 December, and then decreased from 2010 to 2012.
We also show the time profile of reduced-$\chi^2$ ($\chi^2/{d.o.f.}$) in the fit to the PAMELA data on a solar rotation period basis in Figure ~\ref{fg:chi1}.
We find that the reduced-$\chi^2$ in all the fit is much smaller than 1.
It indicates that our calculated spectra agree remarkably well with observations.
Figure ~\ref{fg:chi1} also indicates that $\chi^2$ after 2011 has larger values than those before 2011.

\subsection{Modulation in the polarity reversal period}\label{sec:reversal}
Although the calculated proton spectra show a good agreement with the PAMELA observations in the period 2006/07--2012/10, there are large discrepancies in the subsequent polarity reversal period. The polarity reversal often occurs near the solar maximum. 
	
It is a challenge to model the modulation effect in polarity reversal period.
The gradual reversal process and the frequent solar events disturb the interplanetary medium, therefore the magnetic filed structure becomes more complex in this period.
The diffusion and drift coefficients related to the Parker magnetic field model might not be appropriate during the polarity reversal period.
To account for these fact, we modify the diffusion coefficient by introducing a power law rigidity dependence.
The diffusion coefficient is described as $k_{\perp} \propto k_{\parallel} 
\propto R^\delta $ (DC $\propto R^{\delta}$) for the particle rigidity above $0.1 \, \GV$.
We take $\delta$ as a free parameter rather than a constant 1 in our default case. 
Some studies argue that the drift effect vanishes during the solar maximum \cite{1993AdSpR..13..239P,2001SSRv...97..295P}.
We attempt to turn off the drift effect in our model.

\begin{figure}[!hbt]
   \centering
   \includegraphics[width=.9\textwidth]{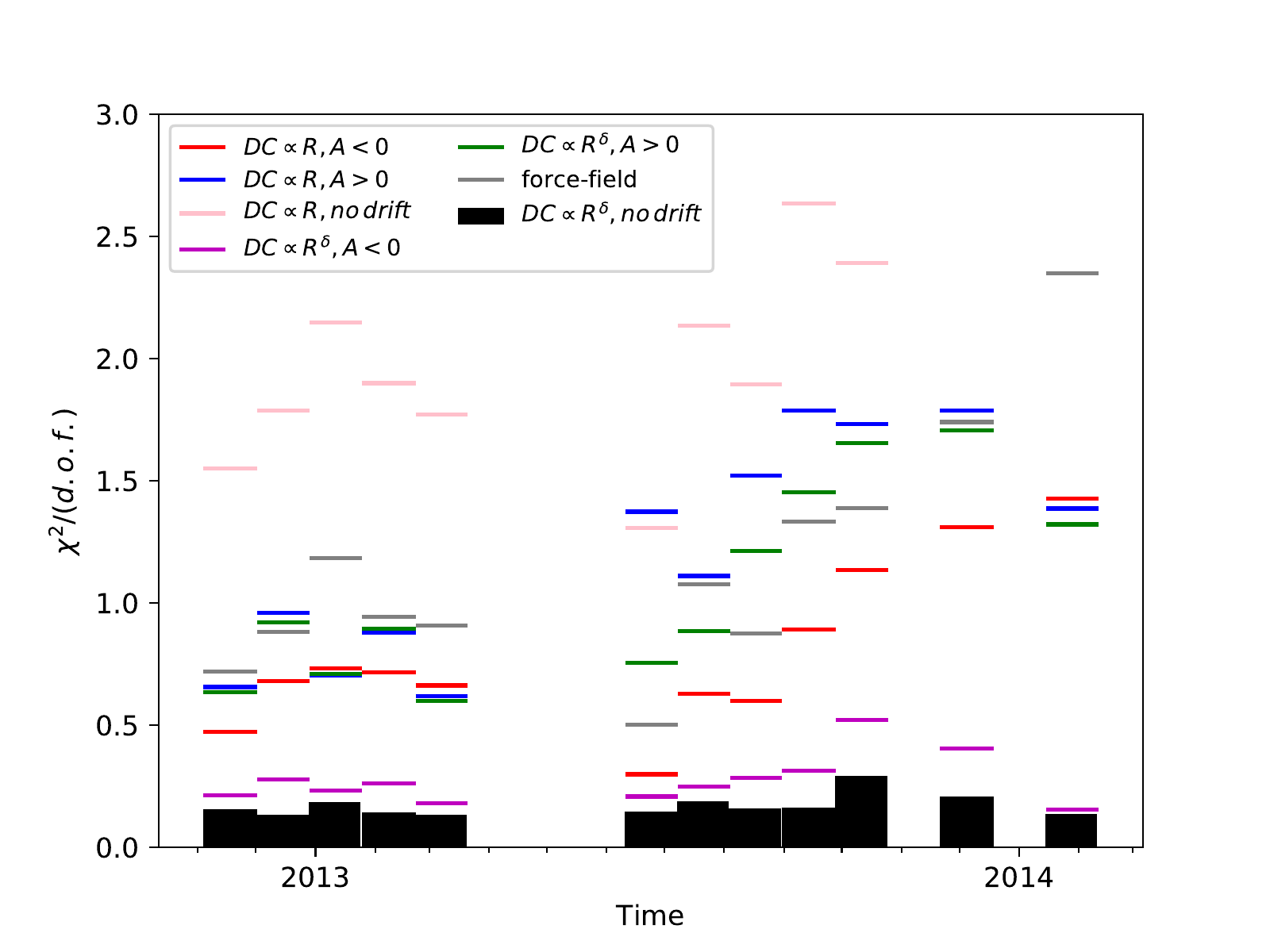}
   \caption{$\chi^2$ from the fits to the PAMELA proton data in the polarity reversal periods for the force-field approximation model and six assumptions on the diffusion coefficient and drift effect.}
	\label{fg:ps}
\end{figure}

\begin{figure}[!hbt]
	\centering
	\includegraphics[width=0.9\textwidth]{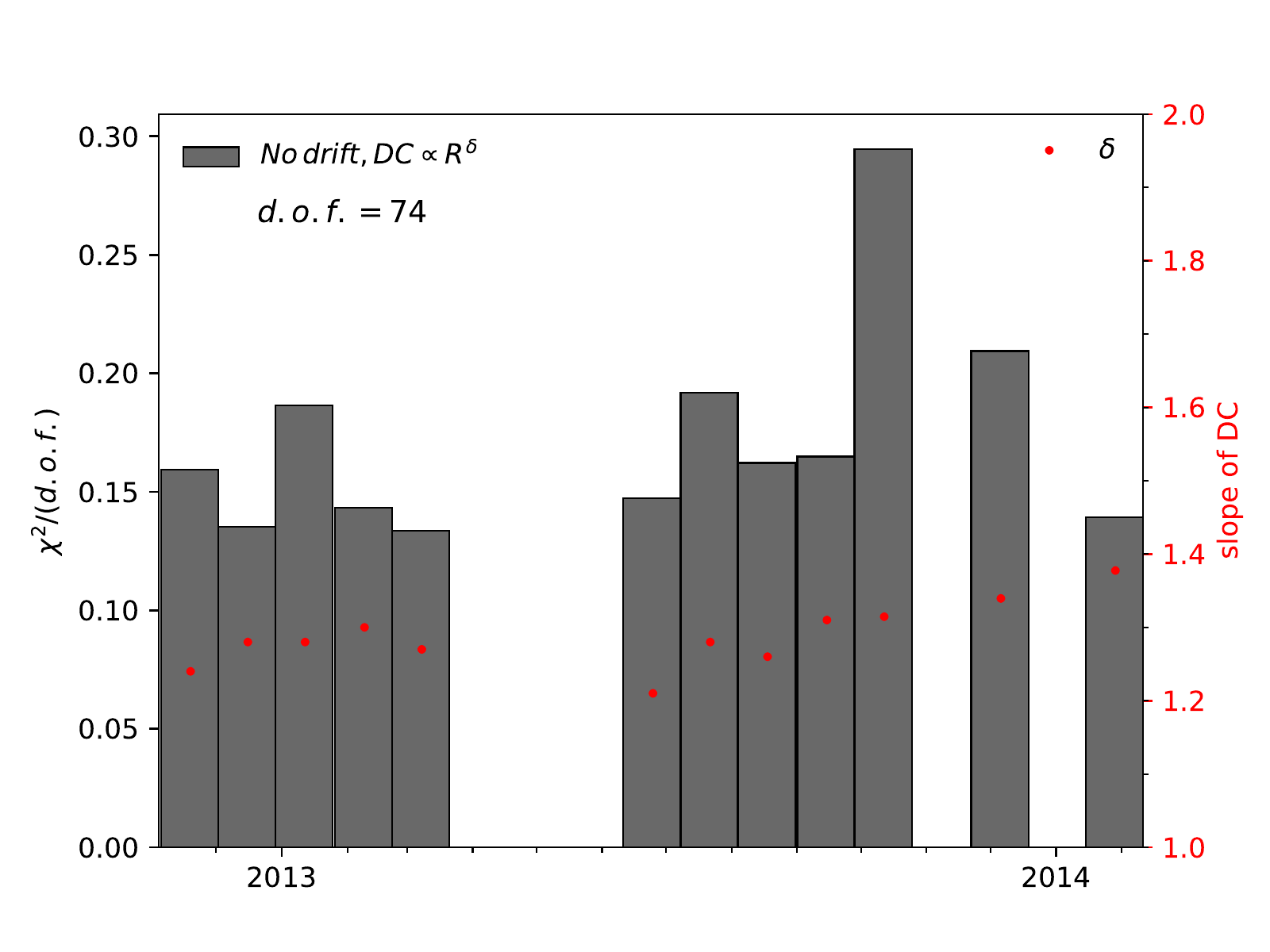}
	\caption{The time profile of reduced-$\chi^2$ and the slope of diffusion coefficient $\delta$ from the fit to the PAMELA proton data during the polarity reversal periods.}
	\label{fg:chi2}
\end{figure}

We attempt some assumptions for the modulation effect in the polarity reversal periods. There are two assumptions for the rigidity dependence of the diffusion coefficient: $DC \propto R$ and $DC\propto R^\delta$. For the drift effect, we consider three cases: the polarity is positive ($A>0$), the polarity is negative ($A<0$) and no drift effect.
Thus there are total six assumptions for the diffusion coefficient and drift effect. 
The force-field approximation model is also included in our comparison.
In Figure \ref{fg:ps} we show the resulting $\chi^2$ from the fits to the PAMELA proton data for different scenarios.
We find that the assumption with a $DC\propto R^\delta$ and without drift provides the best fit. 
It is evident that adopting a variable power law rigidity dependence of the diffusion coefficient can significantly reduce the $\chi^2$. 
The time profile of $\delta$ and reduced-$\chi^2$ for the best fit are shown in Figure \ref{fg:chi2}. 
As the diffusion coefficient is related to the magnetic field power spectrum, the property of the HMF turbulence during polarity reversal should be different in the quiet epoches.

\FloatBarrier

\subsection{An empirical relation between the diffusion coefficient and the magnitude of HMF at the Earth}
\begin{figure}[!hbt]
   \centering
   \includegraphics[width=1\textwidth]{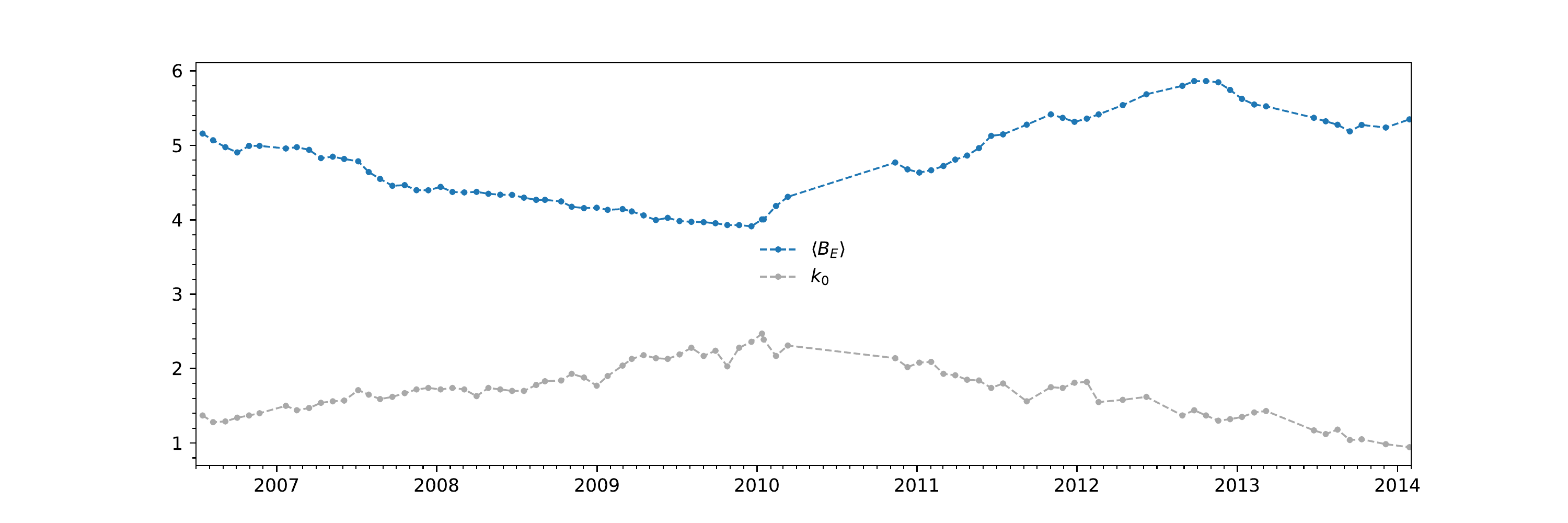}
   \caption{The time profile of $k_0$ and the backward average of the HMF strength $\langle B_E \rangle$.}
   \label{fg:kk}
\end{figure}

It is well known that the diffusion coefficient is anti-correlated with HMF strength.
In this work, the time dependence of the diffusion coefficients is described by a scale factor $k_0(t)$.
The time variations in magnetic field strength are introduced by averaging $B_E$ over the time taken the solar wind to reach the modulation boundary from the Sun. 
Figure \ref{fg:kk} shows the time profile of $k_0(t)$ and backward time average of HMF at the Earth $\langle B_E \rangle$.

\begin{figure}[!hbt]
	\centering
	\includegraphics[width=1\textwidth]{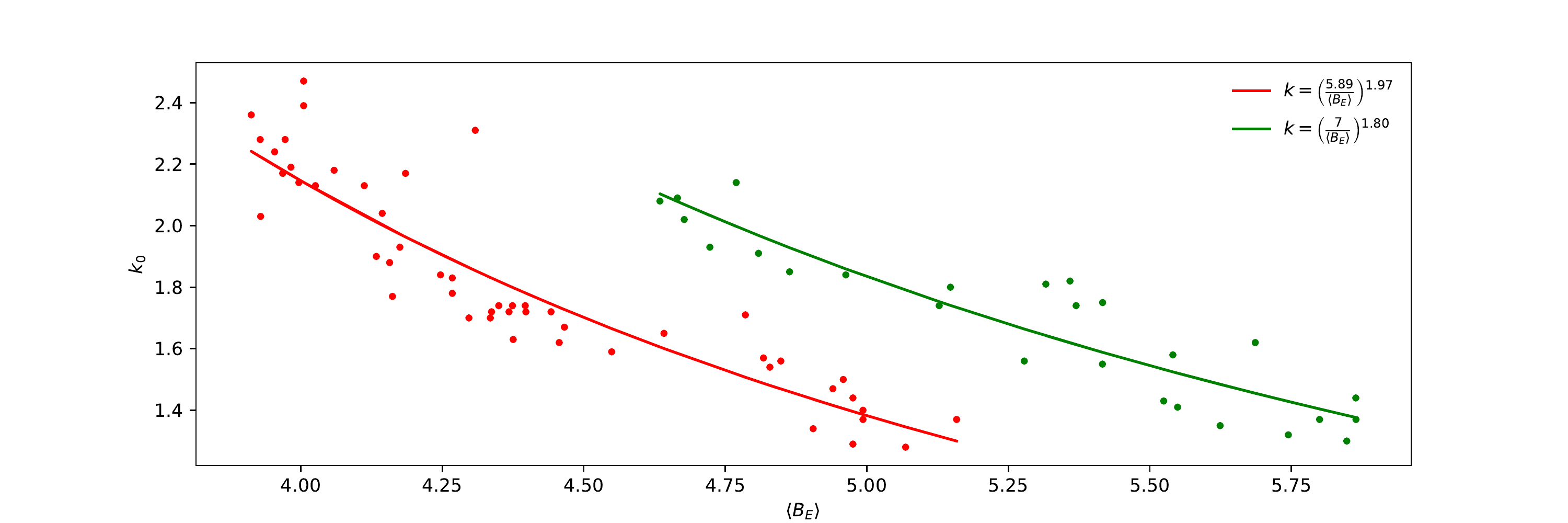}
	\caption{The power law relation between the diffusion coefficient $k_0$ and average HMF strength at the Earth ${\langle B_E \rangle}$. The red and green lines represent the fitting results during 2006/07 to 2010/03 and during 2010/10 to 2013/02, respectively.}
	\label{fg:BKline}
\end{figure}

We follow the previous works \cite{2001AdSpR..27..481P,2004ApJ...603..744F} and assume $k_0 =\left( \frac{B_c}{\langle B_E \rangle} \right)^{n}$, where $B_c$ and $n$ are free parameters.
Figure \ref{fg:BKline} shows the relation between $k_0$ and ${\langle 
B_E \rangle}$ during 2006/07 to 2013/02.
For the period of 2013/06--2014/02, the correlation between $k_0$ and ${\langle B_E \rangle}$ is weak (see Figure 
\ref{fg:kk}), so we do not take into account this period.
Obviously there is a discrepancy between $B_c$ in the declining phase 
(2006/07--2010/03) and increasing phase (2010/10--2013/02) of observed cosmic ray intensity level, while we find that the power $n$ approximates 2 and slightly varies with time. This result is consistent with the conclusion in 
\cite{2002JGRA..107.1353W}.
From the empirical relation found in the fit, we can use the estimated magnetic field strength to obtain the diffusion coefficient and get predictions for the modulated spectra.

\section{LIS for helium, carbon and boron}\label{sec:nuclei}
In order to derive the LIS for other nuclei, we investigate the modulation effect for the CR helium, boron, and carbon.
Here, we also use the cubic spline interpolation method to construct the LIS, and derive the helium LIS by minimizing the weighted differences between the calculated spectra and the observations of Voyager 1 and BESS-POLARII.
The same method is used but the combination of data from the Voyager 1, PAMELA and ACE boron (carbon) observation in 2009 to obtain the boron (carbon) LIS.
The parameters of the LIS are summarized in Table ~\ref{tab:LIS2}.

\begin{table}[!htb]
	\caption{The parameterization of LIS spectrum for helium, boron and carbon with cubic spline interpolation method.}\label{tab:LIS2}
	\centering
	\begin{tabular}{lcccccccl}\hline \hline
	
		&$log(R/\GV)$	 				&-1.0		    &-0.50		&0			&0.50		&1.00		&1.50			&2.00 \\
		He	&$log(I/(\GeV\m^2\s\,\sr))$	&2.4742  	 &2.7412	   &2.6561  &1.7645  &0.4564  &-0.8996   &-2.2882		\\
		B	&$log(I/(\GeV\m^2\s\,\sr))$	&-1.5451     &-0.6089   &-0.3095 &-0.7162 &-3.5311 &-5.0564    &-6.6083		\\
		C	&$log(I/(\GeV\m^2\s\,\sr))$	&-0.7080     &0.2617    &0.4494  &-0.3155 &-1.5787 &-2.9170     &-4.2698	\\
	\hline
	\end{tabular}
\end{table}

\begin{figure}[!hbt]
	\centering
	\includegraphics[width=.9\textwidth]{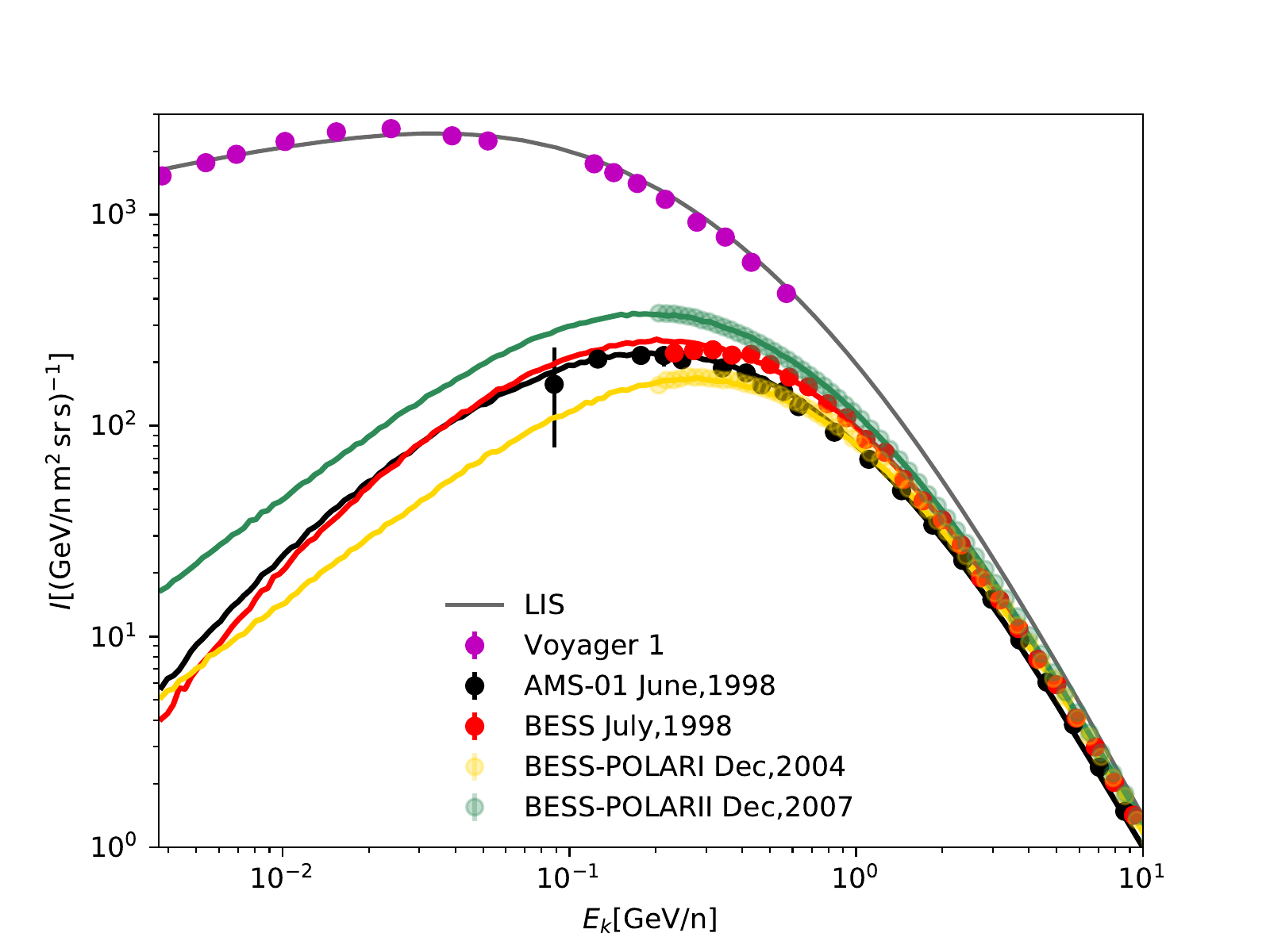}
	\caption{The calculated helium spectra compared to the experimental observations, including Voyager 1, AMS-01 and BESS results.}
	\label{fg:he}
\end{figure}

\begin{figure}[!hbt]
	\centering
	\includegraphics[width=.9\textwidth]{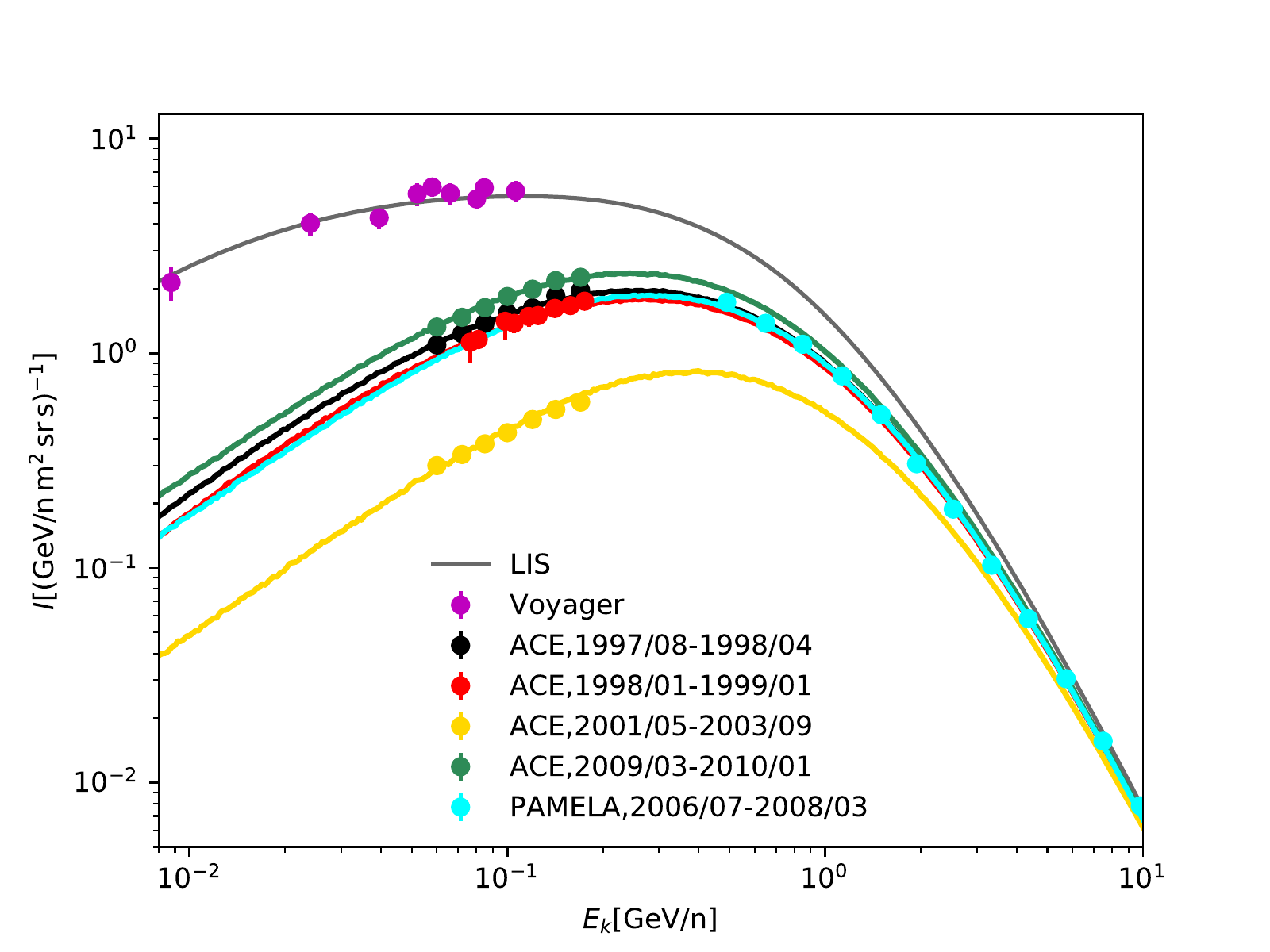}
	\caption{The calculated boron spectra compared to the experimental observations, including Voyager 1, ACE and PAMELA results. }
	\label{fg:boron}
\end{figure}

\begin{figure}[!hbt]
	\centering
	\includegraphics[width=.9\textwidth]{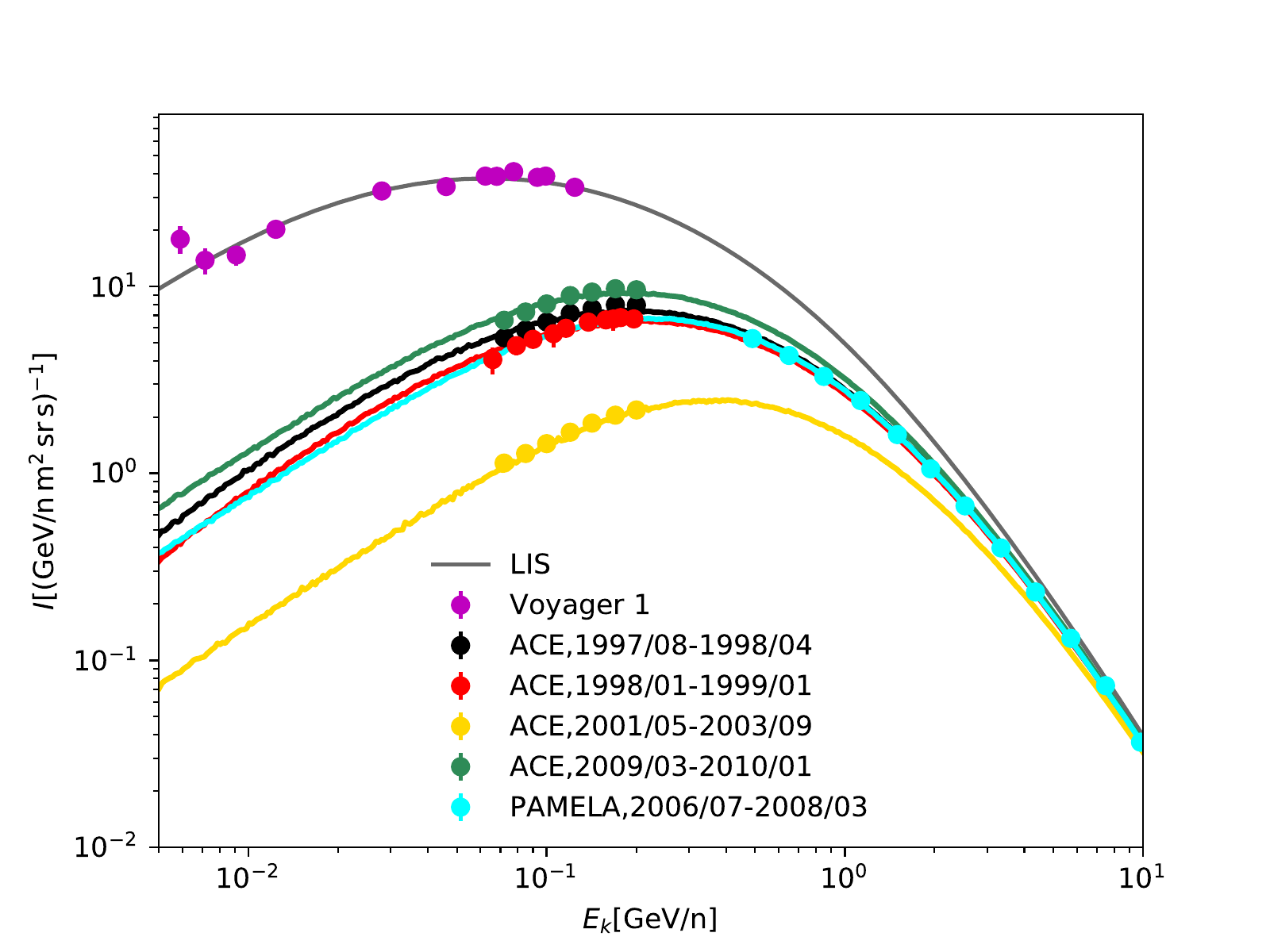}
	\caption{The calculated carbon spectra compared to the experimental observations, including Voyager 1, ACE and PAMELA results. }
	\label{fg:carbon}
\end{figure}
 
In Figure ~\ref{fg:he} we show the calculated CR helium spectra and compare them with the AMS-01, BESS98, BESS-POLARI, and BESS-POLARII results \cite{2000PhLB..494..193A,2007APh....28..154S,2016ApJ...822...65A}. 
The LIS of helium and the Voyager 1 data are also shown.
In order to reproduce the spectra observed by AMS-01, a factor 1.25 is adopted to scale down the LIS.
Figure ~\ref{fg:boron} and Figure ~\ref{fg:carbon} show the LIS and calculated spectra for boron and carbon, respectively. For comparison, the ACE and PAMELA results \cite{2006AdSpR..38.1558D,2013ApJ...770..117L,2014ApJ...791...93A} are also shown.
It is shown that our calculated spectra can well reproduce the observations.
Note that in our calculation, the CR spectra for different nuclei in the same period share modulation parameters. This is an important improvement compared with the force field approximation model, where the potential parameter should be specified for each CR specie \cite{2013ApJ...770..117L}.

\section{conclusion}\label{sec:conclusion}

In this work, we study the solar modulation of CRs and derive the new CR LIS with a time-dependent modulation model and the latest CR experiments. 
The parameter describing the global characteristics of heliospheric environment, such as the solar wind speed,magnetic field magnitude and tilt angle, are all obtained from observations. 
In out default calculation, the only free parameter is the scale factor of the diffusion coefficient.

We adopt the long-term PAMELA observation to derive the LIS and investigate the modulation effect of the CR proton. 
We utilize some data samples of PAMELA during the low solar activity periods and the result of Voyager 1 to obtain the proton LIS, then all the PAMELA data before the polarity reversal can be well reproduced.
Modeling the modulation effect during the polarity reversal period is challenging, since the theory of CR propagation in the heliosphere during this period is poorly understood.
The complex magnetic field configuration increases the uncertainties of diffusion and drift effects. 
In order to reproduce the observations, we change the linear relation between diffusion coefficient and rigidity to $DC \propto R^{\delta}$ and assume there is no significant drift effect in the polarity reversal period.
We find that the diffusion coefficient is anti-correlated to the HMF strength at the Earth. 
An empirical relation can be described as $DC \propto {\langle B_E \rangle}^{-n}$, where $n$ is $\sim 2$ and slightly varies with time.

We also study the modulation effect and derive the LIS for the CR helium, boron and carbon. In the calculation, the parameters in the modulation model are taken to be same for different CR species in the same period. 
Since the calculated spectra can well explain several experimental results, our approach is a good description for dealing with the modulation effect. 
Using the LIS derived here, uncertainties in the study of the CR propagation in the Galaxy can be reduced.

\section*{Acknowledgement}
This work is supported by the National Key R\&D Program of China (No. 2016YFA0400200),
the National Natural Science Foundation of China (Nos. U1738209 and 11851303).
\bibliography{paper_pam3}
\bibliographystyle{h-physrev}
\newpage

\appendix

\section{Comparison with the Force-Field approximation}
\begin{figure}[!hbt]
   \centering
   \includegraphics[width=.8\textwidth]{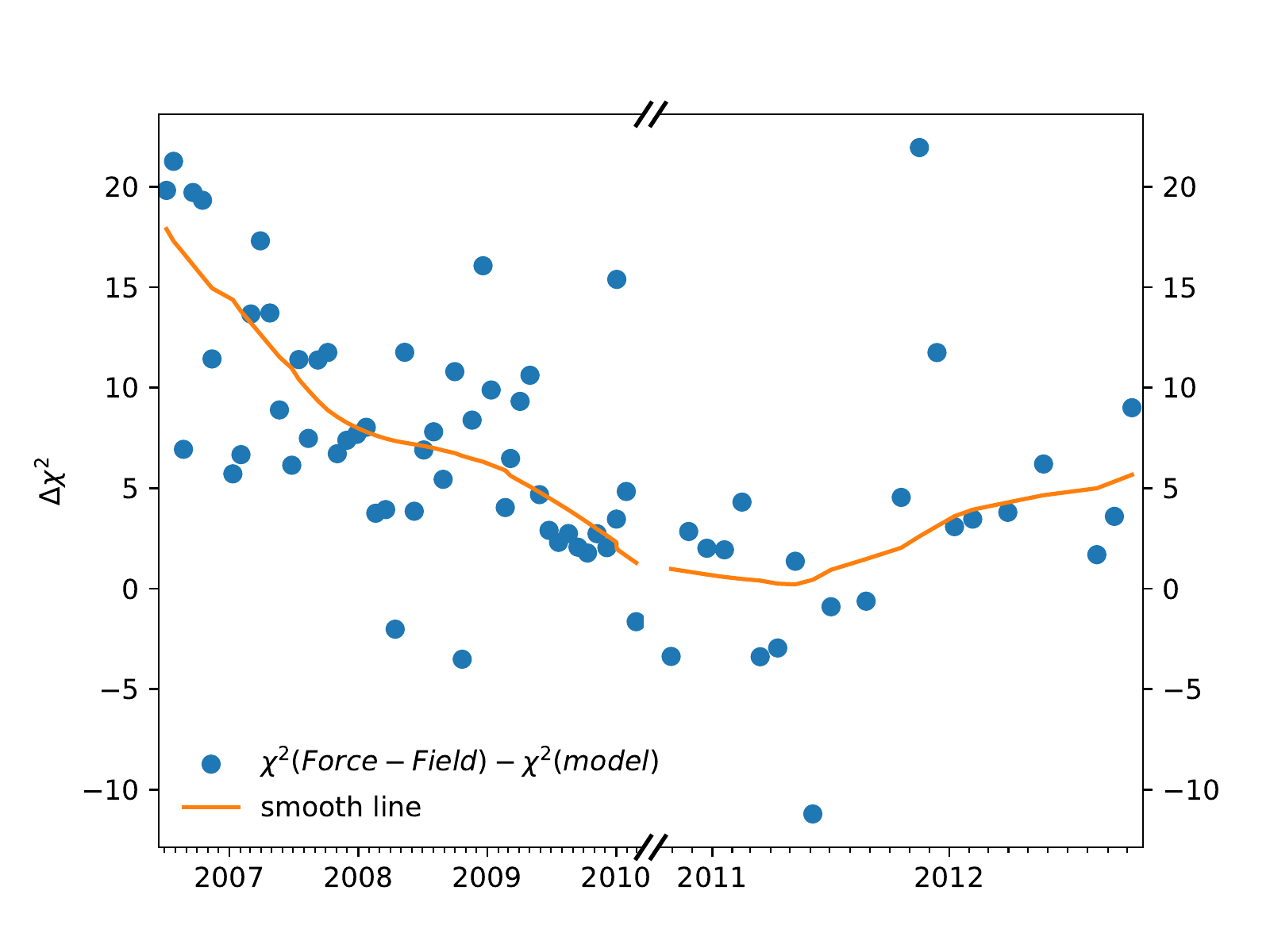}
   \caption{Comparison of $\chi^2$ from the fits to the PAMELA proton data in two models. The dot indicates the difference of $\chi^2$ between our results and the Force-Field approximations. The orange line is the smooth result for the scatter points using the Python statsmodels library \url{https://www.statsmodels.org} with LOWESS (LOcally WEighted Scatterplot Smoothing) method.}
	\label{fg:ffvsmodel}
\end{figure}

In order to compare our results with the Force-Field approximation, we show the difference of $\chi^2$ from the fits to the long-term PAMELA proton observations in two scenarios in Figure \ref{fg:ffvsmodel}. We can see that in most periods the Force-Field approximation give a larger $\chi^2$ than our results.

\end{document}